\RequirePackage[hyphens]{url} 

\documentclass[conference]{IEEEtran}
\IEEEoverridecommandlockouts
\usepackage[switch]{lineno}
\usepackage{amsmath,amsfonts}
\usepackage{graphicx}
\usepackage{textcomp}
\usepackage{xcolor}
\usepackage{hyperref} 
\usepackage{subcaption}
\usepackage{listings}
\usepackage{comment}
\lstdefinelanguage{go}{
	morekeywords=[1]{%
		break,default,func,interface,select,case,defer,go,map},
	morekeywords=[2]{%
		close, write, read},
	morekeywords=[3]{},
	morekeywords=[4]{true,false,iota,nil},
	morestring=[b]{"},
	morestring=[b]{'},
	morestring=[b]{`},
	comment=[l]{//},
	morecomment=[s]{/*}{*/},
	sensitive=true
}
\newcommand\textbfr[1]{\textcolor{red}{\textbf{#1}}}

\usepackage[textsize=small, disable]{todonotes}\newcommand{\todone}[1]{\todo[color=green!40, disable]{#1}} 

\usepackage[inline]{enumitem}

\usepackage[numbers]{natbib}
\makeatletter
\def\thesubsubsectiondis{\thesubsectiondis\arabic{subsubsection}}
\def\subsubsection{%
	\@startsection
	{subsubsection}                 
	{3}                             
	{\parindent}                    
	{3.5ex plus 1.5ex minus 1.5ex}  
	{0.7ex plus .5ex minus 0ex}     
	{\normalfont\normalsize\itshape}
}
\makeatother

\begin{document}
\title{CRISP: Confidentiality, Rollback, and Integrity Storage Protection for Confidential Cloud-Native Computing\\
}

\author{
	\IEEEauthorblockN{Ardhi Putra Pratama Hartono}
	\IEEEauthorblockA{\textit{TU Dresden}\\
		}
	\and
	\IEEEauthorblockN{Andrey Brito}
	\IEEEauthorblockA{\textit{Universidade Federal de Campina Grande}\\
		}
	\and
	\IEEEauthorblockN{Christof Fetzer}
	\IEEEauthorblockA{\textit{TU Dresden}\\
		}
}


\maketitle

\begin{abstract}
Trusted execution environments (TEEs) protect the integrity and confidentiality of running code and its associated data. Nevertheless, TEEs' integrity protection does not extend to the state saved on disk. Furthermore, modern cloud-native applications heavily rely on orchestration (e.g., through systems such as Kubernetes) and, thus, have their services frequently restarted. During restarts, attackers can revert the state of confidential services to a previous version that may aid their malicious intent. This paper presents CRISP, a rollback protection mechanism that uses an existing runtime for Intel SGX and transparently prevents rollback. Our approach can constrain the attack window to a fixed and short period or give developers the tools to avoid the vulnerability window altogether. Finally, experiments show that applying CRISP in a critical stateful cloud-native application may incur a resource increase but only a minor performance penalty.
\end{abstract}


\maketitle
\section{Introduction}

Guaranteeing confidentiality and integrity of data and applications is a traditional challenge in computer science. Nevertheless, recent trends made this quest even more relevant: Digital transformation moves in-person and paper procedures to remote digital platforms~\cite{nguyen:2019trasnformation}; Edge computing moves services and data from well-protected data centers to smaller edge clusters managed by third parties~\cite{edge:2019}; IoT collects detailed sensitive data that is processed remotely to generate valuable insights on industrial applications~\cite{iot-cloud:2020}.

Confidential computing is a powerful tool to help protect data and code processed in remote cloud and edge infrastructures~\cite{iot-cloud:2020}. Tools for confidential computing can provide a confidentiality- and integrity-protected environment for services during execution, 
complementing the general good practices for encrypting data in transit~\cite{tls} and at rest~\cite{StrongBox:2018, BesFS:2020}. Nevertheless, while these Trusted Execution Environments (TEEs) protect memory's integrity and freshness during the execution, the state persisted to storage is not protected off-the-shelf and can still be rolled back if developers do not apply explicit mitigations~\cite{rollback:2017, ROTE:2017, Kumar:securefs}. Unfortunately, stateful cloud-native applications rely heavily on orchestration, having frequent container restart and rescaling operations.  During such a restart, attackers can replace the current state with a previous version~\cite{Ahmadvand:2018:IntegrityProt}.

Interestingly, even if the attacker can only revert the state without being able to modify or even read its content, several powerful attacks can provide good leverage to their malicious goals. For example, cloud operators may access sensitive data and then roll back audit systems to hide their actions. In another example, an adversary could circumvent access revocations and undo undesired actions (e.g., through a repudiation attack) with the same approach. In general, reverting recent undesired changes can be a powerful attack instrument even when the data is confidentiality- and integrity-protected. Our goal is to indirectly protect a stateful confidential application by protecting its state, especially against rollback attacks. 


This work extends tools for confidential computing to enable transparent rollback protection on services with existing integrity and confidentiality protection. With CRISP, we leverage an existing runtime to execute unmodified applications with Intel SGX enclaves. We combine the runtime with a monotonic counter to embed counter values with write operations from the runtime, ensuring rollback protection. Although binding a monotonic counter with a state is not particularly novel~\cite{speicher:2019, sgxlog:2017, enclavedb:2018, cure:2021}, we add multiple distinctive mechanisms while exposing interfaces to mitigate performance losses also  with a tunable parameter.
CRISP can acknowledge operations more optimistically or pessimistically, thereby offering developers additional design-space options. Our generic interface allows for a controllable trade-off between performance and potential vulnerability windows.
We show that having both controllable- and known-vulnerability windows can be easily mitigated by external approaches from both client and server.


Ensuring a confidential application has access to trusted storage is not trivial. Related works were able to provide confidentiality and basic integrity protection to storage such as \textit{Protected Files} in Graphene~\cite{graphene:2017, graphene:pf} and \textit{File System Protection File} in SCONE~\cite{Arnautov:scone}. Unfortunately, both store the metadata in the filesystem, which is vulnerable to rollback attacks. 
Some approaches can withstand a rollback attack by involving a trusted remote party over the network, at the cost of significant performance degradation~\cite{Gregor:cas, Kumar:securefs, nimble:2023}, or through specialized storage~\cite{Krahn:pesos, Ahn:diskshield, cure:2021}. Our approach uses the regular filesystem and can work with different sources for a monotonic counter, such as ROTE~\cite{ROTE:2017}, ADAM-CS~\cite{adamcs:2021}, or LCM~\cite{rollback:2017}. 

Our approach is compatible with the community version of the SCONE runtime. We also evaluated it with an identity provisioning system for zero-trust architectures, SPIRE (SPIFFE Runtime Environment)~\cite{spire:book}. SPIRE uses a local soft state and a persistent state on an SQL backend (SQLite, PostgreSQL, or MySQL). Combining our approach with a highly available database such as MariaDB shows that our protected SPIRE incurs minimal performance degradation.
\todo{Emphasize the novelty}
The contributions of this paper are the following: 
\begin{enumerate}
	\item We detail an approach for tunable rollback-protected storage for confidential stateful services, helping to ensure confidentiality, integrity, and freshness 
	of the data.
	\item Our approach is incremental and equipped with multiple levels of involvement. Applications initially use CRISP transparently but can later incorporate checks to avoid vulnerability windows. We also discuss possible optimizations and their consequences.
	\item We show that our approach incurs minimal performance overhead, evaluating it with microbenchmarks and experiments with realistic workloads, confirming its practicality. 
\end{enumerate}

We organize the rest of the paper as follows. Section~\ref{sec:background} presents background information on trusted execution environments, monotonic counters, and the considered threat model for confidential computing. Sections~\ref{sec:arch} and~\ref{sec:implementation} discuss CRISP's architecture and its implementation, respectively. Section~\ref{sec:eval} presents our experimental evaluation. Sections~\ref{sec:relatedwork} and~\ref{sec:conclusion} present related work and some concluding remarks.





\section{Background and Threat Model}\label{sec:background}

\subsection{Trusted Execution Environments, Intel SGX, and SCONE}

Using Trusted Execution Environments (TEEs) is one approach to enable trusted computing, where computations are protected from malicious actors even if they compromise infrastructure components. One of the most popular approaches to such an environment is Intel SGX~\cite{intel:sgx}. With SGX, processes can create enclaves, segregated regions of memory that are accessible only to the enclave that created it. This approach protects the enclave memory from other applications in the same host and even from components with higher privileges, such as the operating system and the hypervisor~\cite{sgxexplained:2016}.

Unfortunately, the minimization of the trusted computing base that enables secure enclaves also comes with a cost: the operating system is untrusted, and, therefore, enclaves cannot perform system calls~\cite{sgxexplained:2016}. Such limitation required software redesign so that data could be fed to and collected from the enclaves. To address the limitation, the community proposed several approaches. Some argue that a minimum operating system should be part of the enclave, such as a library OS (e.g., Graphene~\cite{graphene:2017}). Others argue that a mediator should handle the system calls and ensure the operating system cannot compromise operations (e.g., SCONE~\cite{Arnautov:scone}). In this case, the mediator can block or transform risky operations. 

We build our work atop SCONE as it has a community version that offers several features we can leverage. Legacy applications run in SCONE by being recompiled with the SCONE compiler. 
When starting a SCONE application, an enclave is created and subsequently communicates with a remote Configuration and Attestation Service (CAS) server, providing information about itself, which then receives secure configurations~\cite{Gregor:cas}. In a trusted environment, such configurations include application inputs, environment variables, and application secrets.

The SCONE runtime supports data confidentiality and integrity through the file system shield~\cite{scone:docs}, known as FSPF. The runtime encrypts all read and write operations transparently to the application. To ensure the integrity of the files, SCONE uses a Merkle tree. The root hash is named the \textit{tag} and is checked or updated inside the enclave. When a program starts, the runtime retrieves the expected tag from CAS. When a program finishes or does some disk synchronization (\textit{fsync}, \textit{fdatasync}, or similar mechanism), it also updates the tag with its CAS server~\cite{Gregor:cas, scone:docs}. Therefore, this mechanism can be used to detect initial filesystem integrity and could be used to detect rollbacks if the runtime sends all writes to the remote CAS. Unfortunately, sending the tag to CAS on every disk synchronization is expensive.

SCONE offers a feature called \textit{vault file} to address the performance cost of communicating with a remote CAS. When using this feature, the SCONE application performs the initial attestation and obtains configurations from the remote CAS, but afterward stores them in the vault file. The application can then read the configurations and filesystem expected tag from the local vault file as needed. In addition, when using the vault file, the application can store the updated file system tags on the local storage, improving performance. However, since all data to restart the application is on the local storage, rolling back the state of an application only requires the attacker to replace the state of the local filesystem with a previous version, which also includes the previous version of the vault file.

\subsection{Monotonic Counters}

A monotonic counter is a mechanism often used to prevent rollback attacks in trusted systems~\cite{adamcs:2021, ROTE:2017, Trinc:2009, Ariadne:2016, speicher:2019}. Since a monotonic counter is incremented but not decremented or reset, systems can use the counter to track process evolution.

Monotonic counters are available through different implementations. Intel SGX has made monotonic counters available inside enclaves for some processors. However, their performance and usage limits rendered it impractical~\cite{ROTE:2017}. Trusted Platform Modules (TPM) also provide monotonic counters, which, unfortunately, also suffer the same performance issues, being rather short-lived~\cite{ROTE:2017, Ariadne:2016}. The Embedded Multi Media Card (eMMC) standard, starting from version 4.4, introduced Replay Protected Memory Block (RPMB)~\cite{rpmb:2015},  which stores a secure-monotonic counter~\cite{StrongBox:2018}. 

Table~\ref{tab:mcperf} shows a comparison between the monotonic counter implementations. Although we do not require a specific implementation, we adopt an RPMB-based F-Secure USB Armory~\cite{usbarmory} as our monotonic counter model. The \textit{emmc-RPMB} has a much longer life, improving maintainability, and lower latency when compared to the other commercial-off-the-shelf (COTS) products. Our aim is to keep the stack as common and straightforward as possible. Although monotonic counter services such as ROTE~\cite{ROTE:2017} and \mbox{ADAM-CS~\cite{adamcs:2021}} offer higher performance, their complex implementation hinders wide adaptation.


\begin{table}[]
	\caption{\label{tab:mcperf} Performance of considered monotonic counter implementations}
	\centering
	\begin{tabular}{|l|l|l|l|}
		\hline
		\textbf{Implementation} & \textbf{Usage limit} & \textbf{Write latency} & \textbf{Read latency} \\ \hline
		SGX~\cite{ROTE:2017} & 1.05 million & $80-250\ ms$ & $60-140\ ms$ \\ \hline
		TPM~\cite{Ariadne:2016} & 300k - 1.4m & $\approx 25ms$ & $\approx 15\ ms$ \\ \hline
		4 ROTE~\cite{ROTE:2017} servers & unlimited & $1-3\ ms$ & $\approx 15\ ms$ \\ \hline
		emmc-RPMB & 2\textsuperscript{32}-1 & $19.97\ ms$ & $3.8\ ms$ \\ \hline
	\end{tabular}
\end{table}

\subsection{Threat Model} \label{sec:threat}
Our work builds on Intel SGX. We, therefore, start with Intel SGX's threat model. This model states that the processor and application code are bug-free. In addition, the attacker resources do not enable them to physically inspect the internals of a running processor or break currently-recommended cryptographic algorithms. Side-channel and availability attacks are out of scope. The adversary aims to compromise the integrity and confidentiality of data and code and has full administrative access to the hardware and software stack.

In addition to the threats mentioned above, we consider attacks against storage. Specifically, we focus on ensuring data CIF (Confidentiality, Integrity, and Freshness), especially rollback protection to files in storage volumes. Adversaries can also gain complete storage control, including taking and restoring old snapshots and having multiple replicas. 

We assume the adversary cannot roll back or tamper with the monotonic counter. We assume the provisioning of the monotonic counter (e.g., RPMB's one-time key generation) is fully trusted (within enclaves or controlled by the operator of the secure application). Replacing the counter hardware can, therefore, be detected. Another important assumption is access to a trusted time source. Detailing such implementation is out of the scope of this paper. A simple solution is to have one or more enclaves that access a trusted remote time server and track time locally. SGX version 2 enclaves have access to the TSC register, which is trusted as long as the enclave does not lose the CPU. Because they detect when they get evicted from the CPU, enclaves can track and serve time and resynchronize with external sources if all lose CPU simultaneously.

Although our threat model does not include side channels and availability issues, they are of practical value. Thus, we consider mitigations to them. More specifically, we consider a layered security approach where enclaves run in hardened (virtual or physical) machines that use cured signed images and measured boots (e.g., assisted by a TPM), or even runtime machine attestation such as in \cite{triglav}. This makes it more challenging to collocate malicious workloads and secure enclaves. Runtime features such as Varys~\cite{varys} and BROFY~\cite{Hartono2021brofy} also help make side channels even harder to exploit. Then, we mitigate availability attacks by using cloud orchestrators such as Kubernetes. Because service unavailability is visible externally and may also affect the income of cloud providers, all involved parties have the incentive to address them.

\section{Architecture} \label{sec:arch}
To protect against data rollback, we use a trusted Monotonic Counter (MC). In Section~\ref{sec:fspfmc}, we will show that the MC is incremented and included in the state saved on each local disk synchronization. Throughout our explanation, we denote `disk synchronization' as \texttt{fsync}, although other invocations (e.g., \texttt{sync}, \texttt{fdatasync}) are also included. The combination of MC increment, data encryption, and metadata handling makes the data durable and protected against rollback attacks. However, this approach may be impractical in a program that does intensive synchronization. Next, in Section~\ref{sec:batching}, we elaborate on a batching mechanism to counter this issue. Despite introducing an unavailability window, as we disclose in Section~\ref{sec:vuln}, this approach may significantly improve the performance since it reduces the frequency of increments. We then introduce a mechanism for developers to eliminate vulnerability windows.

In summary, we use SCONE runtime to ensure the \textbf{confidentiality} of the application. The data is always encrypted at rest (through SCONE FSPF), during processing (through Intel SGX), and in transit (through SCONE network shield~\cite{scone:docs} or existing protocols such as TLS). Should the data be compromised at rest (\textbf{integrity} violation), the runtime detects it at the program startup thanks to SCONE FSPF and our rollback-protection mechanism. To guarantee \textbf{freshness}, we set up the runtime to \textit{commit} (increment MC and update tag) on two crucial occasions: on each explicit disk synchronization and when closing the file or ending the program. Therefore, data written to the disk is guaranteed to be protected.

\subsection{Transparently Protect Storage} \label{sec:fspfmc}

Our primary approach is incrementing the MC on every flush (e.g., \texttt{fsync}) and referring to it as part of the storage metadata. Using SCONE FSPF, the data at rest will be encrypted. Here, a protection key will be randomly generated and then stored with CAS. The runtime will then store a data chunk in a secure memory before encrypting it and writing it to the disk. Nevertheless, an application can still read it transparently via the runtime, and no modification is needed. In this approach, we update the FSPF tag in three situations: disk flush, file close, and program exit. The latter two are particularly important since both are well-known gateways to rollback attacks.

Since each flush will increment the MC, we bind each tag update to an MC value, all saved in the local vault file. Although pushing the MC value to the SCONE CAS is possible, there is no benefit. Rolling back only the vault file can be easily detected, as the MC value in a rolled-back vault file will be lower than the one on the actual MC. 

Adversaries can attempt to roll back the whole environment: vault file, encrypted data, and the MC. However, tampering with or resetting the MC is difficult and is not included in our threat model. Similarly, replacing the MC with a new one with an equivalent counter is detectable because the initial provisioning ties the device to a key. Tampering with an Intel SGX enclave is also challenging and assumed to be unfeasible in our threat model. 


When a program starts, it loads the vault file and the FSPF volume (metadata). First, the runtime checks the integrity of these files. Specifically, the latest volume tag will be loaded from the vault file and matched with an Merkle tree root hash for the relevant part of the filesystem (e.g., a directory). Then, the runtime will compare the MC value tied with the loaded tag to the one in the MC. If the value in the MC is higher than the stored one, a rollback is suspected, and the runtime halts. Restarting the program is only possible after both the FSPF volume and vault file are either restored to the correct state or removed. When restarting with a new state, a new counter is created in the same device. 

The protection described above is transparent to the application and requires no source code or library modifications. Nevertheless, the MC latency will considerably slow down the write operations. The following section describes an approach to mitigate this issue.


\subsection{Optimistic Batching} \label{sec:batching}

We introduce \textit{optimistic batching}, an approach that combines multiple tag updates from flush operations into a single MC increment. In such a case, the runtime promises an MC value for every tag update but does not immediately increment the MC. We implement this approach by adding a separate loop thread to the runtime portion inside the enclave, named the \texttt{mc-thread}. This loop thread is independent of the application threads and will process the accumulated operations, update the tag once, and increment the MC once. Note that in this approach, we squash all tag updates into a single tag.

\begin{figure}[t]
	\includegraphics[width=1.05\linewidth]{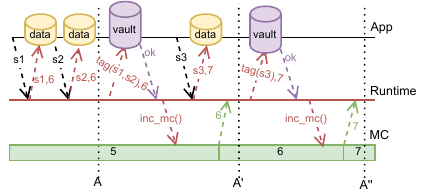}
	\caption{Batching request process}
	\label{fig:opbatch}
\end{figure}

As illustrated in Figure~\ref{fig:opbatch}, we identify flush requests received by the runtime from the application with \texttt{s1}, \texttt{s2}, $\ldots$, \texttt{sN}.  The runtime acknowledges immediately but puts each request into a time-aware queue. Periodically, it consolidates multiple requests, generating a single tag and one MC value. For example, in Figure~\ref{fig:opbatch}, \texttt{s1} and \texttt{s2} are included in batch 6. Any requests after this point (denoted as \texttt{A}) are on the next batch. After confirming the writes to the FSPF volume and vault file, the runtime issues MC increments. Batch 6 is only protected after the runtime receives the acknowledgment from the MC (denoted as \texttt{A'}). Note that at this point, batch 7, which includes \texttt{s3}, is not yet (rollback) protected, even though its contents are already flushed to the disk.

Although flush requests execute asynchronously, other system calls that trigger disk-related synchronization are still done synchronously, namely \texttt{close} and \texttt{exit} system calls. The runtime does not batch these operations as they may represent the end of data processing from the application. Consequently, these calls are blocked until all outstanding asynchronous requests are committed, including vault-file writes and MC increments, guaranteeing a consistent state. 

The introduction of optimism with the batching is especially relevant to enable reasonable throughputs for the slower yet widely available hardware monotonic counters, like those on SGX or TPM. However, the consequence is that an adversary can roll back the updates within a batch. We discuss this vulnerability window in Section~\ref{sec:vuln}. Our following approach reduces transparency to give more control to applications, enabling them to know whether the data they committed to disk (e.g., through \texttt{fsync}) are already rollback protected. 


\subsection{Checker API}
\begin{figure*}
	\centering
	\includegraphics[width=0.9\linewidth]{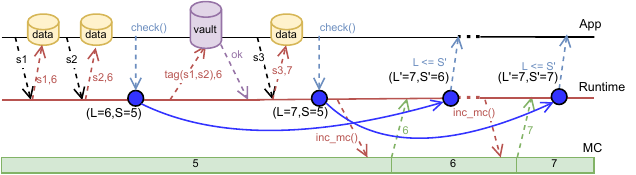}
	\caption{Checker API course}
	\label{fig:checkchan}
\end{figure*}

To complement optimistic batching, we introduce the Checker API to enable applications to ensure that flushed writes are MC-protected. The call to this API is blocking, meaning the runtime is waiting for the confirmation of the asynchronous operations. This API is language-agnostic and is available on a local network connection. The blue lines in Figure~\ref{fig:checkchan} illustrate the following example. Suppose a multithreaded service that calls the Checker API after the return of \texttt{s2}; CRISP holds two internal values in the runtime, \textit{local} and \textit{stable}, represented as \texttt{L} and \texttt{S}. \textit{Local} is the latest value written to the vault file and represents the promised MC value. \textit{Stable} is the latest MC value. 

The Checker API returns only when the current \textit{stable} value is greater or equal to the \textit{local} value at the moment of the call (condition shown as \texttt{L <= S'} in the figure). For example, the first \texttt{check()} call sees 6 and 5 as \textit{local} and \textit{stable}, respectively, and waits until \textit{stable} reaches at least 6.


Developers can combine optimistic batching and call to the Checker API at their discretion, enjoying the flexibility of design-space options. One approach to maintain transparency in the application is to implement calls to the Checker API at a library level. For example, libraries that abstract storage interactions (such as with databases, logging, or filesystems) or that handle communication with external parties can use the Checker API, eliminating the need to implement this in the application. The application operator controls optimism by configuring shorter batch time limits or forcing the Checker API after each flush, which results in synchronous behavior as discussed in Section~\ref{sec:fspfmc}. Alternatively, one could consider all flush-related system calls as we exemplify in Section~\ref{sec:checkerapiimpl} with exposed frequency configuration. Note that a rogue cloud operator cannot change these configurations as they are included in the secure configuration after getting attested.


\subsection{Discussion on Vulnerabilities} \label{sec:vuln}
\todo{Discuss SoCC reviews regarding TSC?}


\smallskip
\noindent{\textbf{Availability:}} If the adversary rollbacks the data, the vault file, or both, the runtime detects and blocks the execution. Also, suppose the application crashes at specific points, for example, after writing a new tag to the vault file but before the increment of the MC. In that case, the program will not be able to start due to the MC value discrepancy. Another unavailability will arise if the application crashes when the data has been written (for example, on \texttt{s3} in Figure~\ref{fig:opbatch}). If the tag is not up-to-date in the vault file (point \texttt{A'}), the application also cannot be started since the hash of the disk state (in this case, \texttt{s3}) will not match against the one in the vault. In both of those cases, a fresh environment is required to be able to execute the program again. Finally, adversaries can hold the MC hostage and slow it down. If the waiting time on the queue exceeds the timeout threshold, the runtime will halt. Such scenarios introduce a tradeoff between integrity and availability. In production use cases, protected systems may need additional replication to cope with the increased failure rates. 

\todo{What are the failure rates in a realistic, stressful scenario?}


\smallskip
\noindent{\textbf{Integrity: }} As discussed in the previous section, combining batching and optimism introduces a \textbf{vulnerability window}. In Figure~\ref{fig:opbatch}, if an attacker wants to roll back the filesystem changes incurred by \texttt{s3}, the vulnerability window starts from the submission of \texttt{s3} to the point when the increment of the MC from 6 to 7 happens internally at the MC. Because the runtime acknowledges writes optimistically (before the MC is available for another increment), the vulnerability window can last twice the maximum write latency, resulting in $40\ ms$ for our RPMB-based MC (see Table~\ref{tab:mcperf}).
For this particular case, we emphasize the relevance of the Checker API. If the service calls the API before externalizing any information that is a consequence of \texttt{s3}, the application will be blocked until the MC is up-to-date (see the second \texttt{check()} in Figure~\ref{fig:checkchan}). 

When the developer adds check calls after critical writes, the service will not externalize any information that needs to be protected against rollbacks. Thus, if the adversary rolls back the system, it will be equivalent to a benign crash-restart, where operation \texttt{s3} may or may not have made it to disk, and the client that triggered the operation may not have assumed an acknowledgment. Since the only writes that can be rolled back did not need to be protected, there is no effective window of vulnerability.

Note that CRISP does not batch the \texttt{fsync} call itself; it batches the call to  the FSPF metadata population and MC increment. Therefore, it doesn't deteriorate \texttt{fsync}-semantics. Figure \ref{fig:opbatch}, label \texttt{A'} describes where data has persisted but is not necessarily protected against rollbacks. Furthermore, with the adversary presence, \texttt{fsync} also does not guarantee the changes are durable since they could roll back changes beyond any \texttt{fsync}, which is prevented by CRISP. 
\todo{Further explanation on CRISP behaviour on this vuln window? Review A SoCC}
%
%


\section{Implementation} \label{sec:implementation}

In our approach, we protect applications without source code modifications. To ease wide adaptation, we deliberately choose commercial off-the-shelf (COTS) MCs with interfaces accessible from within the SCONE runtime. Consequently, the configuration and optimization options can be executed securely and communicated through SCONE CAS~\cite{Gregor:cas}. 

We applied CRISP to MariaDB, a popular database, to validate our approach and guide optimizations. We then evaluate how to extend this protection to SPIRE. SPIRE is the open-source SPIFFE reference implementation ~\cite{spire:book}. SPIFFE is an open-source standard defining identity formats, interfaces, and workflows for automated management of identity provisioning processes~\cite{spiffe}. Both SPIFFE and SPIRE are graduate projects of the Cloud Native Computing Foundation, which are considered stable and proven in production~\cite{cncf}.

%

As with other modern cloud-native applications, container orchestrators can scale and manage SPIRE services. Thus, SPIRE delegates the management of its state to a separate storage, in this case, MariaDB. Because the MariaDB stores mappings between selectors and IDs, it is a critical component for the security of the role cluster and needs to be protected.

\smallskip
\noindent{\textbf{Stack details.}} We do not assume a particular MC implementation. However, we use the data from Table \ref{tab:mcperf}  to simulate an RPMB-based MC on the same machine due to its being a COTS product with a lower adaptation barrier. Our stack uses MariaDB version 10.11.4, SPIRE version 0.12.0, and SCONE Community Edition version 5.7. CRISP's implementation is in the SCONE runtime, keeping the source code of MariaDB and SPIRE unmodified. For this scenario, the developer could include the Checker API usage in \textit{gorm}\footnote{\url{https://gorm.io}} (an ORM library used in SPIRE), in SPIRE
, or in MariaDB itself. 



\subsection{Checker API} \label{sec:checkerapiimpl}
The implementation of the Checker API spawns a new thread for each check request that is received through a TCP connection. First, it will check the latest promised MC value that has been written to the vault file, named as \textit{local}. It will then check the value on the MC periodically (named \textit{stable}) until the condition is satisfied (\textit{stable} is greater or equal to \textit{local}). \textit{Local} is not necessarily bound to the preceding \texttt{fsync} request. Therefore, there might be a case where the callee needs to wait for at most one new increment to be completed. Furthermore, if the queue is empty and there is no pending increment, the Checker API will return immediately. 

CRISP exposes an option to enable internal probabilistic checking out of the box. When enabled, all fsync-related calls will be intercepted, and the blocking Checker API call will be triggered according to the chosen probability. Effectively, this will block the callee’s thread progression. In this approach, no new thread is spawned as there is no TCP connection to handle. This option must be securely enabled as a secret provisioning via SCONE CAS, preventing an adversary from prolonging the batch size.

\lstset{language=Go,
	basicstyle=\ttfamily\small,
	keywordstyle=\color{blue}\ttfamily,
	stringstyle=\color{red}\ttfamily,
	commentstyle=\color{brown}\ttfamily,
	frame=single}

\begin{lstlisting}[float,language=go,caption={Checker API example usage in gorm.},label=lst:gorm,belowskip=-0.8\baselineskip, aboveskip=-0.2\baselineskip]
func cb_checkerAPI(...) {
  connection,_ :=dial("tcp",checkerAPI_address)   
  connection.write(...) //this line will block
  connection.read(...)  //returning MC value
  connection.close()
}

func init() {
  ....
  register_callback("gorm:after_update", 
      cb_checkerAPI)
  ....
}
\end{lstlisting}

\medskip
\noindent{\textbf{Additional design-space options.}} The combination of the known vulnerability window and the Checker API explores additional design-space options. The latter offers great flexibility in where and when it is used. As shown in Listing~\ref{lst:gorm}, one could call the Checker API after every \texttt{update} query in gorm, which could be implemented using a minimally invasive callback mechanism. The Checker API can also be integrated with other framework components. For example,  communication interceptors (e.g., SCONE network shield, service mesh) could prevent the externalization of uncommitted critical values (such as access control changes or certificate renewals). A flexible approach could also be taken on the client side, such as reconfirming selected operations after the known vulnerability window on the server. Ultimately, developers have a wide range of options to iteratively reduce either overhead or vulnerability windows, or discover the balance between both. 

\todo{Discussion on how to counter vuln window (server, client)}
\subsection{Optimization parameters}
CRISP exposes other tuning parameters: \textit{MC rate limit} and \textit{Queue timeout}. 

\smallskip
\noindent{\textbf{\textit{MC Rate Limit}}} aims to prevent the fast exhaustion of the MC. Setting the \textit{MC rate limit} to a value higher than its write latency provides two benefits: reduced overhead and a longer lifetime of the MC hardware. For example, setting a $100\ ms$ value leads to increased batch sizes, reducing tags generated and encryption tasks. The security impact is the same as using a slower MC: a larger vulnerability window. 

\smallskip
\noindent{\textbf{\textit{Queue timeout}}} regulates the tolerance for MC increment latencies and aims to prevent attackers from slowing down MC operations to increment the window of vulnerability. It is tuned according to the write latency of the MC implementation at hand and is especially relevant when the Checker API is not used. When a request waits in the queue longer than the timeout, the runtime will exit prematurely.  

\medskip
\noindent{\textbf{Default parameter settings.}} We decided on various parameters related to our approach based on a few initial microbenchmarks. We set the number of \textit{enclave threads} (application threads available inside the enclave) depending on the application, namely $4$ for multithreaded and $1$ for single-threaded applications. When not explicitly mentioned, MC Rate Limit and queue timeout optimization are disabled. If the MC Rate Limit is enabled, the queue timeout will add the rate-limiting delay to the actual MC timeout. 


\section{Evaluation} \label{sec:eval}
%
We consider four evaluation scenarios. First, we show raw disk benchmarks on a single-threaded application. Second, we evaluate the impact of calling the Checker API on performance and vulnerability windows. Then, we show how CRISP-equipped MariaDB fares on a benchmark suite. Lastly, we show the performance in production-like experiments with SPIRE, including in a distributed SPIRE scenario. We performed all experiments in servers equipped with Icelake-SP processors (experimental versions that precede the recently released Xeon Scalable 3rd Generation). 

Throughout the experiments, we refer to \textit{Native} and \textit{HW} as the runs executed in a default native environment and on top of the SCONE runtime with Intel SGX hardware protection, respectively. \textit{HW+FSPF} adds SCONE's filesystem protection (FSPF encryption). \textit{MC} and \textit{MC-Optimized} are based on CRISP without and with optimization, respectively. The optimization adheres to SCONE's recommendations for improving IO responsiveness of IO-bound loads by exercising spinning before blocking on a system call queue. In an optimized variant, we also set the  \textit{MC Rate Limit} to $100\ ms$ in order to improve the MC's lifetime to a reasonable extent.

For MariaDB, we enable MariaDB table encryption in the native version and when using SCONE without FSPF. We also set the page size to $4096$ bytes and enabled 1-page caching to increment MC more often, protecting the state as fine-grained as possible. When run on top of a trusted environment, the memory heap is set to $4\ GB$. We left other MariaDB configurations as default.

\subsection{Basic evaluation: Raw disk experiment} \label{sec:expbasic}


Here, we experimented with a simple disk benchmark. We limit the file size to $256\ MB$ and vary the buffer size, with a maximum size of 32768 bytes according to \textit{POSIX} standards. We call the \texttt{write()} system call to write the buffer to an opened file. Then, \texttt{fsync()} is immediately called afterward. As for the reading, \texttt{read()} system call is called with the identical buffer size as writing. 

The writing performance with an MC variant is faster than the native, except for the largest buffer size. This improvement is due to the caching mechanism when using both FSPF and optimistic batching. The performance is slightly worse with an increased MC rate limit to $100\ ms$, as depicted by \textit{HW+Rate Limit}. When CRISP is active, the runtime will cache all write operations and flush them to a memory-mapped IO when a threshold is reached, which, in this case, is close to the \texttt{SSIZE\_MAX} from \texttt{limits.h}, defined by POSIX. Therefore, the runtime ensures that the file and the cache stay synchronized in multiple chunks. The advantage is two-fold: reducing the context switch caused by system calls and hiding the encryption cost from end-users since it encrypts the whole chunk on a single flush.  
\todo{discuss mmap-"async" fsync}
\todone{add MC+Rate limit only. And discuss. queue timeout is not in exp since it's only a threashold/flag}
\begin{figure}[h]
	\centering
	\includegraphics[width=0.7\linewidth]{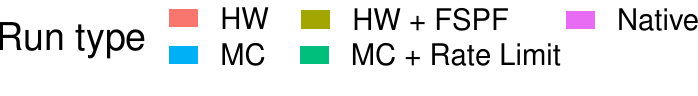}
	\begin{subfigure}[t]{0.48\linewidth}
		\centering
		\includegraphics[width=\linewidth]{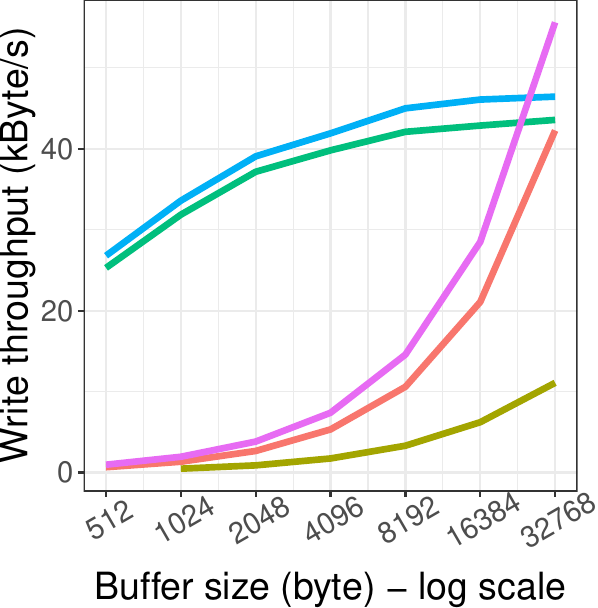}
		\caption{Raw writing}
		\label{fig:rawwrite}
	\end{subfigure}
	~
	\begin{subfigure}[t]{0.48\linewidth}
		\centering
		\includegraphics[width=\linewidth]{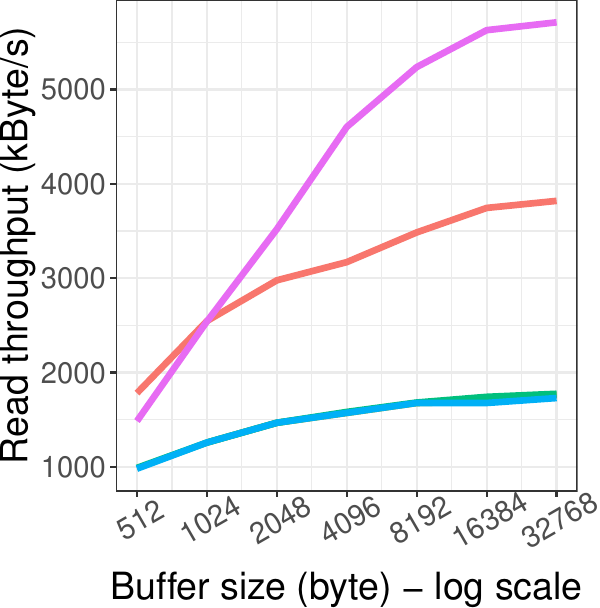}
		\caption{Raw Read}
		\label{fig:rawread}
	\end{subfigure}
	\caption{Raw disk performance experiment}
\end{figure}	

Figure~\ref{fig:rawwrite} depicts the write overhead throughput across multiple execution types. As expected, the \textit{HW+FSPF} version performs poorly since it needs to perform the encryption, update the tag, and write to three different files on every flush (the actual data file on disk, the vault metadata file, and the FSPF metadata). Meanwhile, the \textit{HW} version only has a slight overhead due to system call handling in SCONE for Intel SGX.

\todo{Can we cache read? What's the limit on mmap? related to SCONEHEAP?}
Although CRISP's caching usage consistently masks write overheads, we cannot see the same effect when reading. Figure~\ref{fig:rawread} illustrates the results of reading the file sequentially. A single read operation in native translates to three reads in all variants with FSPF (actual data file on disk, vault metadata, and FSPF metadata), including CRISP. On top of this, there is also the FSPF decryption. This phenomenon does not occur in the HW version, which does not use file protection and, hence, shows no slowdown. 

\todone{But both can read ahead, right? -> \textit{not sure. I think fspf-variant fetch the whole block regardless it's being read right now/not.}}

\subsection{Examining the Checker API} \label{sec:expchecker}
This section assesses various aspects of Checker API usage, including its impact on vulnerability windows, its performance when batching is disregarded, and its feasibility of conscious use.

\begin{figure}[b]
	\centering
	\includegraphics[width=0.9\linewidth]{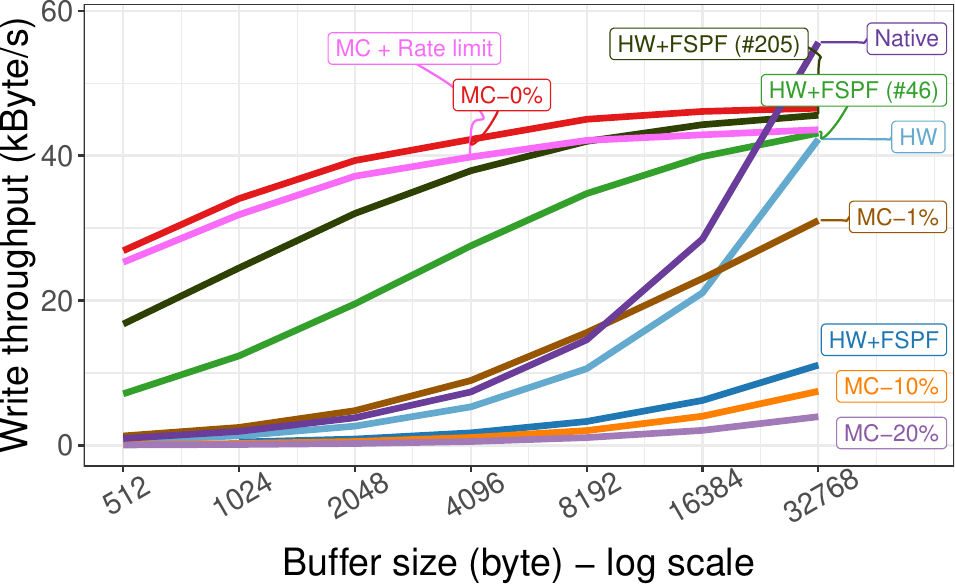}
	\caption{Raw write throughput with various configurations} 
	\label{fig:rawcheck}
\end{figure}

\smallskip
\noindent{\textbf{Checker API and vulnerability window size}.} Here, we evaluate to what extent the length of vulnerability windows can be reduced as well as the impact on performance. Figure~\ref{fig:rawcheck} illustrates the effect of the Checker API in the previous raw disk experiment. Following the prototype specified in Section~\ref{sec:checkerapiimpl}, the callee is implemented internally in the runtime.  Note that for practical usage of the Checker API, the developer should place the checker calls just after the relevant operations (e.g., could skip logging operations). 

We set the probability of having the Checker API called to $1\%$, $10\%$, and $20\%$, in which, representing \textit{MC-1\%}, \textit{MC-10\%}, and \textit{MC-20\%} in Figure \ref{fig:rawcheck}, respectively. \textit{MC-0\%} is a variant that does not invoke the Checker API (0\% probability).

First, we note that invoking the Checker API on a single-threaded disk-intensive application heavily impacts throughput. With $1\%$ checker probability, the average relative throughput compared to native and \textit{MC-0\%} is $77\%$ and $31\%$, respectively. Increasing the probability to $10\%$ is enough to bring it below the HW+FSPF variant. 

Table \ref{tab:rawcheck} illustrates the effect of internally invoked Checker API on various probabilities. Two batching metrics per incremented MC are of interest: size, represented by the number of disk synchronizations (\texttt{fsync}) for a single tag update, and duration, which denotes for the time taken to confirm a single batch protection. On a single-threaded application, calling the Checker API lessens the contention between the \texttt{mc-thread} and the main thread  for a write access to the file system. It occasionally pauses the main thread, letting the \texttt{mc-thread} to promptly proceed with the subsequent batch. This results in a smaller batch size and shorter duration. 

Calling the Checker API more often also decreases the number of disk synchronization (\texttt{fsync}) on a single batch. Since the window of vulnerability is min-bounded by the MC's latency, by having fewer \texttt{fsync} calls, the attack window is reduced. In our experiment, having a $1\%$ probability already considerably improves the length of the vulnerability window, despite having only a $31\%$ average throughput compared to the MC variant on a single-threaded case. Note also that without optimistic batching, the number of \texttt{fsync} is exactly one, while the duration is at the minimum equal to MC's latency. Furthermore, in our experiment, as the check probability increases, the duration's variance is negligible.



\begin{table}[t]
	\caption{\label{tab:rawcheck}Calling Checker API effect on batching metrics (vulnerability window length) - single-thread application.}
	\begin{tabular}{cll}
		\hline \textbf{Check prob. (\%)} & \textbf{avg. \#fsync} & \textbf{avg. duration (ms)} \\ \hline
		0       & 205.97       		& 63.49 \\
		1       & 46.99 (22.8\%)   & 33.528 (52.8\%)\\
		10      & 7.30 (3.5\%)    & 34.158(53.7\%)\\
		20      & 3.922 (1.9\%)    & 35.269 (55.5\%)
	\end{tabular}
\end{table}

\medskip
\noindent{\textbf{CRISP's performance with no batching}.} For completeness, we also add a comparison with an imaginary case where HW+FSPF is batching requests with the size of 205 and 46, inspired by with $1\%$ checks call and without, respectively (See Table~\ref{tab:rawcheck}, 2\textsuperscript{nd} column). MC variant without the call is $15\%$ faster than HW with FSPF (batch size 205), shown as \textit{HW+FSPF (\#205)} in Figure~\ref{fig:rawcheck}. Meanwhile, MC with $1\%$ call is $47\%$ slower than HW with FSPF (batch size 46), shown as \textit{HW+FSPF (\#46)}. Here we show that our optimistic batching performs well, even if we disregard the batching itself. However, our internal checks approach is suboptimal, especially in this particular worst-case scenario (single-threaded and IO-intensive). Placing the checks properly could significantly improve the performance, although it is case-dependent.

\medskip
\noindent{\textbf{Conscious use of Checker API on multithreaded setting}.} For the second experiment, we evaluate the impact of a conscious use of the Checker API, where the developer knows that some of the write operations do not need to be rollback protected (e.g., statistics logging). We implemented a thread pool-based webservice in Rust. Each request will be served by exactly one of these operations: 
\begin{enumerate*}[label=\textbf{(\#\arabic*)}]
	\item Read a file, then return;
	\item Read a file, write to a file, call fsync, then return; And,
	\item read a file, write to a file, call fsync, call the Checker API, then return.
\end{enumerate*}
Thus, operations \#1 and \#2 do not trigger the Checker API, simulating non-critical operations. 
We specify operation~\#2 to serve 20\% of the requests. We vary the threadpool size to match the number of parallel connections in the Apache HTTP Benchmarking tool (\texttt{ab})\footnote{\url{https://httpd.apache.org/docs/2.4/programs/ab.html}}. We also vary operation~\#3 likelihood and then observe the overall average latency on each run as well as its batching metrics. We ran this experiment 500 times with 1000 requests each.

\begin{figure}[ht]
	\centering
	\includegraphics[width=0.8\linewidth]{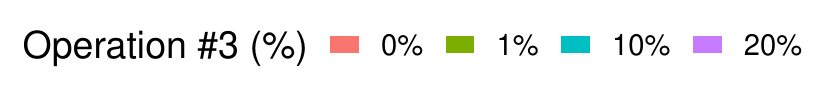}
	\begin{subfigure}[t]{0.45\linewidth}
		\includegraphics[width=\linewidth]{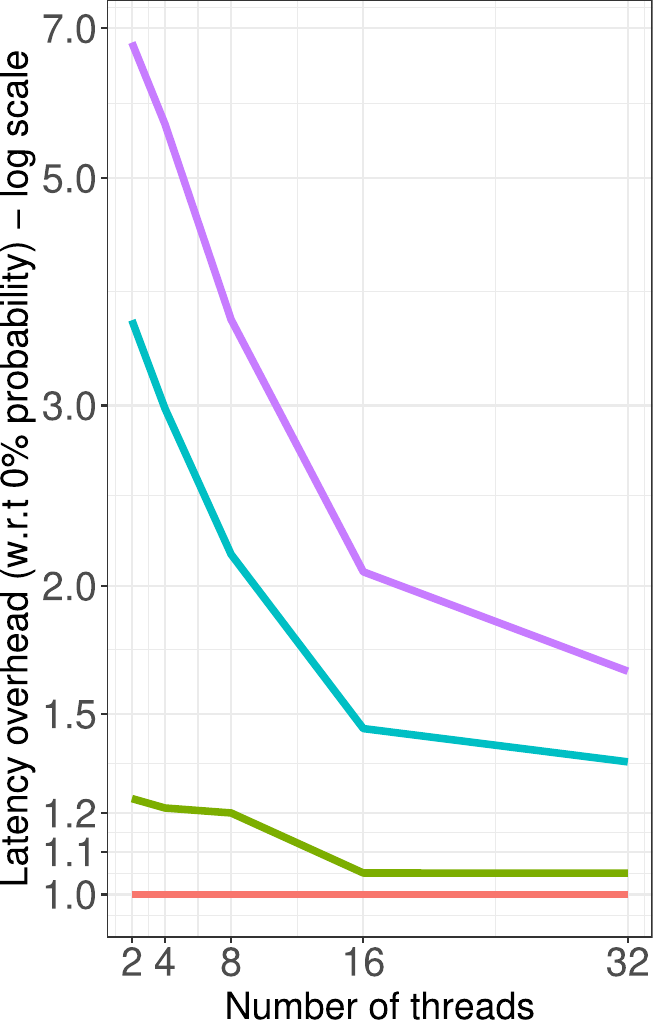}
		\caption{Avg. latency overhead}
		\label{fig:webcheck-perf}
	\end{subfigure}
	~\hspace{5pt}%
	\begin{subfigure}[t]{0.5\linewidth}
		\includegraphics[width=\linewidth]{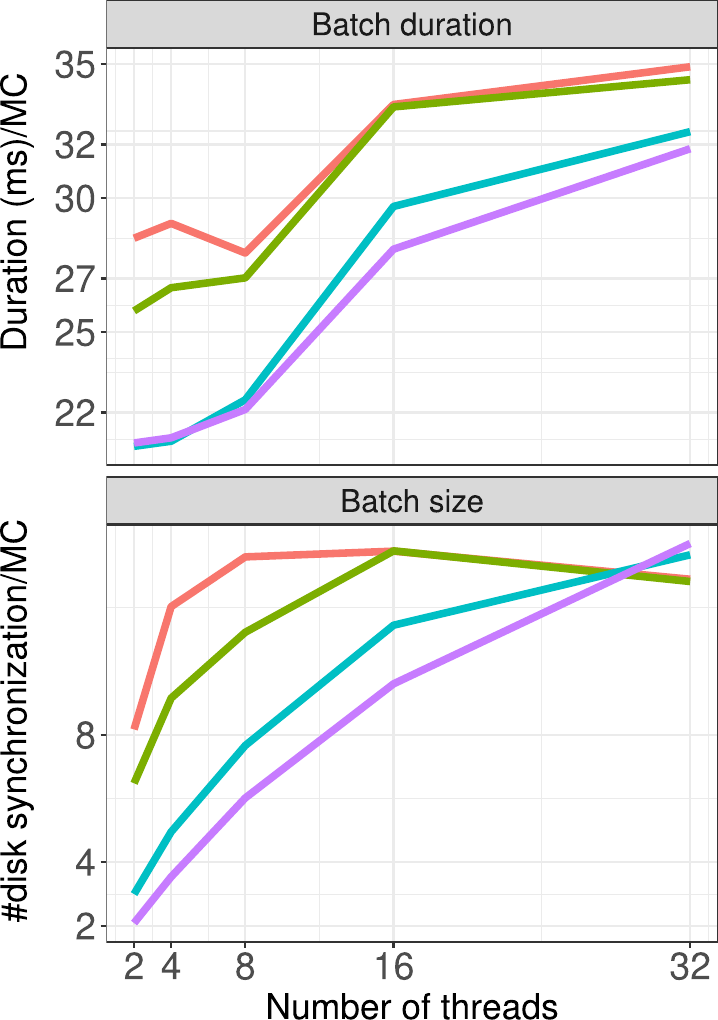}
		\caption{Batching metrics}
		\label{fig:webcheck-vuln}
	\end{subfigure}
	\caption{Checker API-enabled webservice experiment on multithreaded settings}
\end{figure}

As expected, consciously calling the Checker API in multithreaded settings thrives better compared to in a single-threaded setting, especially when it is done more often. In a single-thread program, calling Checker API means that no other operations could continue since it is a blocking operation. Multithreaded programs are able to alleviate this issue and result in lower latency, given that not all threads check the API simultaneously. As depicted in Figure \ref{fig:webcheck-perf}, on 32 threads, the overall latency overhead is $4\%$, $34\%$, and $65\%$ for operation~\#3 being $1\%$, $10\%$, and $20\%$ of all the operations, respectively. 

Regarding the batching metrics, the multithreaded application has shown the same behavior as the single-threaded one: more checking means shorter overall duration and smaller batch size. Higher thread usage, however, causes the impact on calling the API to be less significant. With two threads, the \texttt{fsync} frequency and duration without invoking the API are $3.3\times$ and $1.2\times$ higher compared to operation~\#3 being  20\% (half of all writes), respectively. Meanwhile, with 32 threads, it is only $0.9\times$ and $1.04\times$. Figure \ref{fig:webcheck-vuln} portrays this phenomenon as the thread number increases. 

Increasing the threads will increase the number of possible operations: Multiple threads lead to higher utilization (less idle waiting) and, hence, more \texttt{fsync} operations on a particular time window. Thread contention and IO bottleneck also make the duration longer, leading to more pending operations. Nevertheless, suppose the Checker API was used for all critical operations. In that case, this is not a vulnerability window as no return will be provided for a relevant request, and rolling back the system is equivalent to a benign crash, as we discussed in Section~\ref{sec:vuln}.


\subsection{TPC-C Benchmark on CRISP-enabled MariaDB} \label{sec:expdb}
Next, we experiment with a realistic IO-intensive application, namely MariaDB. In particular, we run a TPC-C benchmark~\cite{tpcc:1993}, which aims to emulate an e-commerce system. It measures not only IO operations but also the computation power of a database system by involving complex database transactions, expressing the results in transactions per minute (\textit{tpmC}). We vary the number of warehouses and connections to simulate a higher level of complexity on the server and increase the load of simultaneous requests.

\begin{figure}[t]
	\hspace*{-0.25cm}
	\centering
	\includegraphics[width=1.001\linewidth]{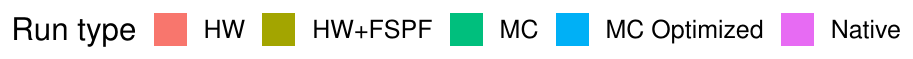}
	\hspace*{-0.15cm}
	\begin{subfigure}[t]{0.48\linewidth}
		\centering
		\includegraphics[width=\linewidth]{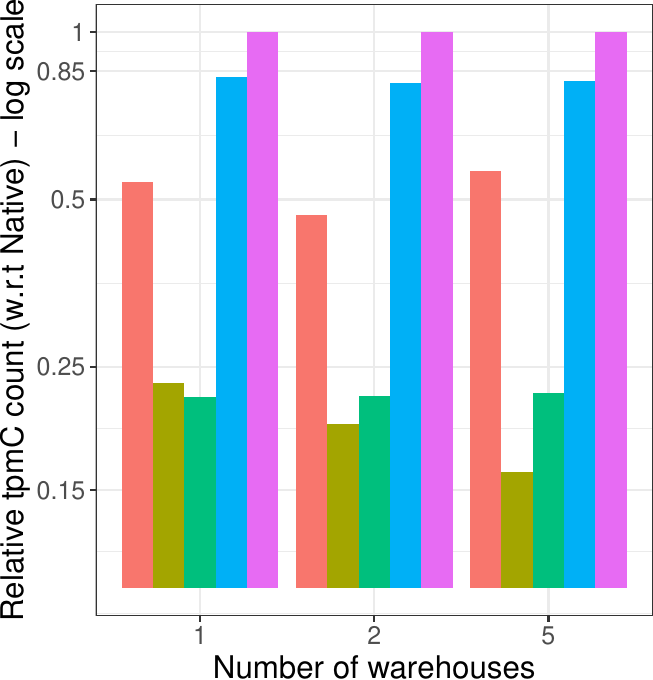}
		\caption{Relative tpmC throughput with varying warehouses}
		\label{fig:mdbwr}
	\end{subfigure}
	~\hspace{5pt}%
	\begin{subfigure}[t]{0.48\linewidth}
		\centering
		\includegraphics[width=\linewidth]{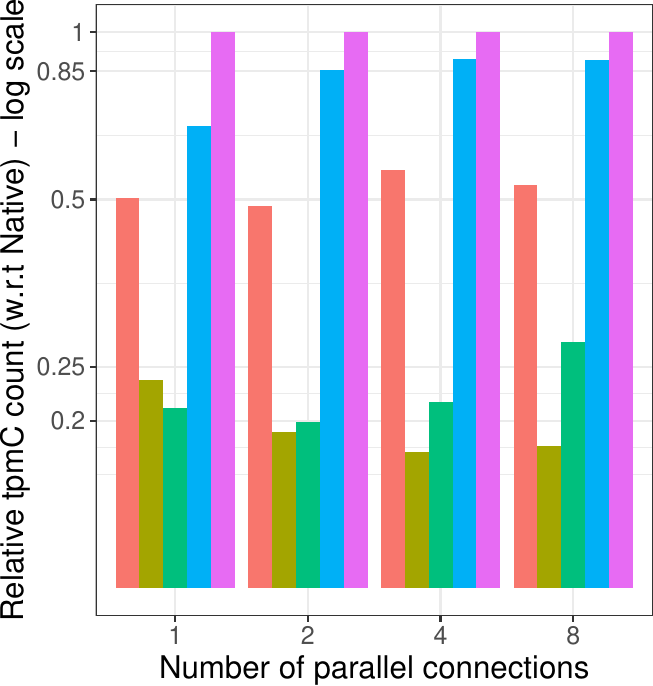}
		\caption{Relative tpmC throughput with varying connections}
		\label{fig:mdbconn}
	\end{subfigure}
	\caption{MariaDB TPC-C experiment}
	\label{fig:mdbtpcc}
\end{figure}

As seen in Figure~\ref{fig:mdbtpcc}, as the number of warehouses and connections increases, CRISP (with \textit{MC Optimized}) is able to keep up with the native version. The non-optimized variant also scales, albeit with lower performance, showing that CRISP's design is compatible with a higher workload. For both \textit{MC} and \textit{MC-Optimized}, the existence of the exclusive \texttt{mc-thread} relieves each connection from waiting for the IO operation to complete since the request will be immediately replied. Moreover, optimistic batching combines several expensive disk-synchronization operations into a single one, improving performance. As a result, the optimized version with CRISP shows $0.83\times$, $1.59\times$, and $3.7\times$ throughput compared to the native, HW, and non-optimized versions, respectively.


Meanwhile, using FSPF dramatically reduces performance and hinders scalability. The \textit{HW} variant, however, is relatively consistent with roughly half of the native's performance. Employing CRISP's approach addresses both issues with more respectable performance and additional protection.


\subsection{Production-like evaluation with SPIRE} \label{sec:expspire}

Here, we observe how a typical application using our approach at the data store will behave. In this case, we deploy SPIRE, which typically consists of the server and one or more agents. The agent performs service (workload) attestation for users, while the server attests agents and acts as a centralized signing authority. The server manages identities and workload entries on a backend (MariaDB, in our case). We do not change any part of the SPIRE code.

We aim to observe day-to-day SPIRE operation, namely the certificate signing throughput. First, we deploy a single SPIRE server and create the entries for the upcoming agents. Then, every 15 minutes, we spawn a new agent for a maximum of 25 instances. Each agent should be responsible for 9000 SVID entries. We set up the SVID's TTL randomly between 20 to 120 seconds. For SVIDs halfway through expiration, agents will send a certificate signing request to the server, adding extra work. The server will then perform the signing of such a request, update the entry on its cache, and send the acknowledgment back. We note the number of certificates the server could successfully sign per second. We also observe the CPU usage on both the server and MariaDB. 

\begin{figure}[t]
	\centering
	\includegraphics[width=1.03\linewidth]{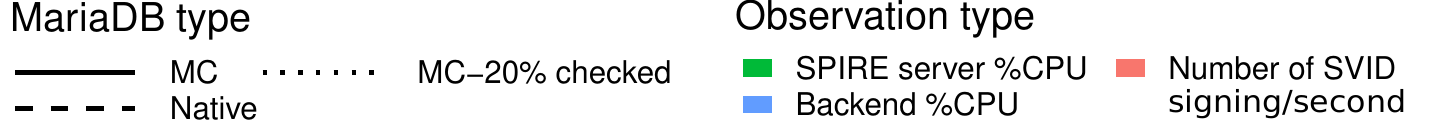}
		%
		\centering
		\includegraphics[width=\linewidth]{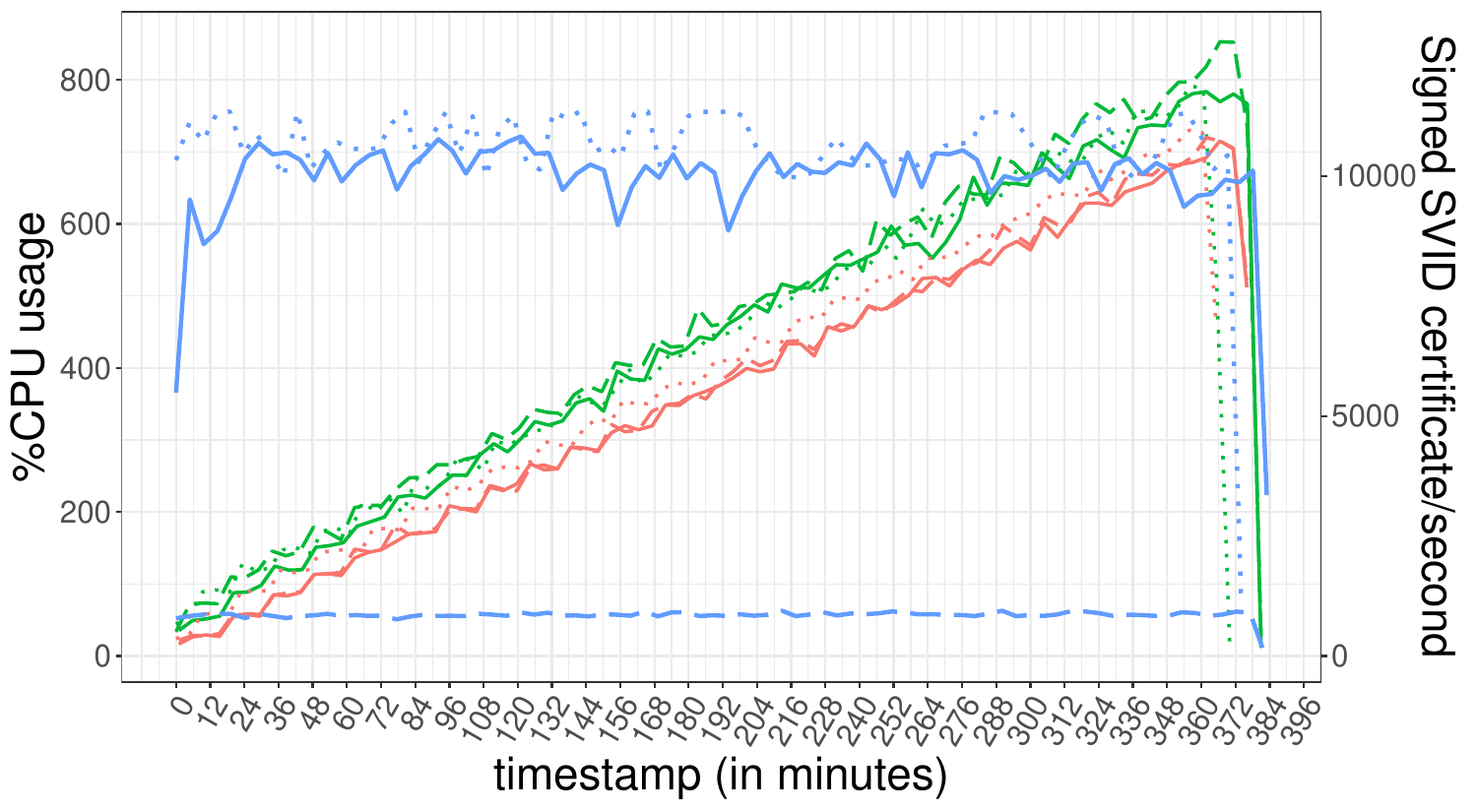}
	\caption{Certificate signing throughput with CPU load on the SPIRE server and MariaDB backend}
	\label{fig:spirewrk-mdb}
\end{figure}	

As depicted in Figure~\ref{fig:spirewrk-mdb}, where both the agents and the server are deployed in a native environment, the throughput constantly increases as the number of active agents increases over time. However, the SCONE configuration favors spinning over blocking to enhance IO performance, leading to high CPU usage from the CRISP-powered MariaDB perspective. Meanwhile, IO usage on the backend (not shown in the graph) is relatively constant and indistinguishable between CRISP and native execution.


We also refer back to the performance penalty caused by calling the Checker API. Similar to in Section~\ref{sec:expchecker}, the callee is implemented internally. Figure~\ref{fig:spirewrk-mdb} shows that despite having a modest check probability ($20\%$), the overhead of SVID certificate signing is negligible. This performance contrasts with our previous experiment in Figure~\ref{fig:rawcheck}, in which $20\%$ checks reduced the throughput to merely $2\%$ on average for the native case. In this experiment, the number of writing operations from the SPIRE server to the CRISP-equipped MariaDB is not intensive; therefore, there is no significant overhead.

\begin{figure}[th]
	\centering
	\includegraphics[width=\linewidth]{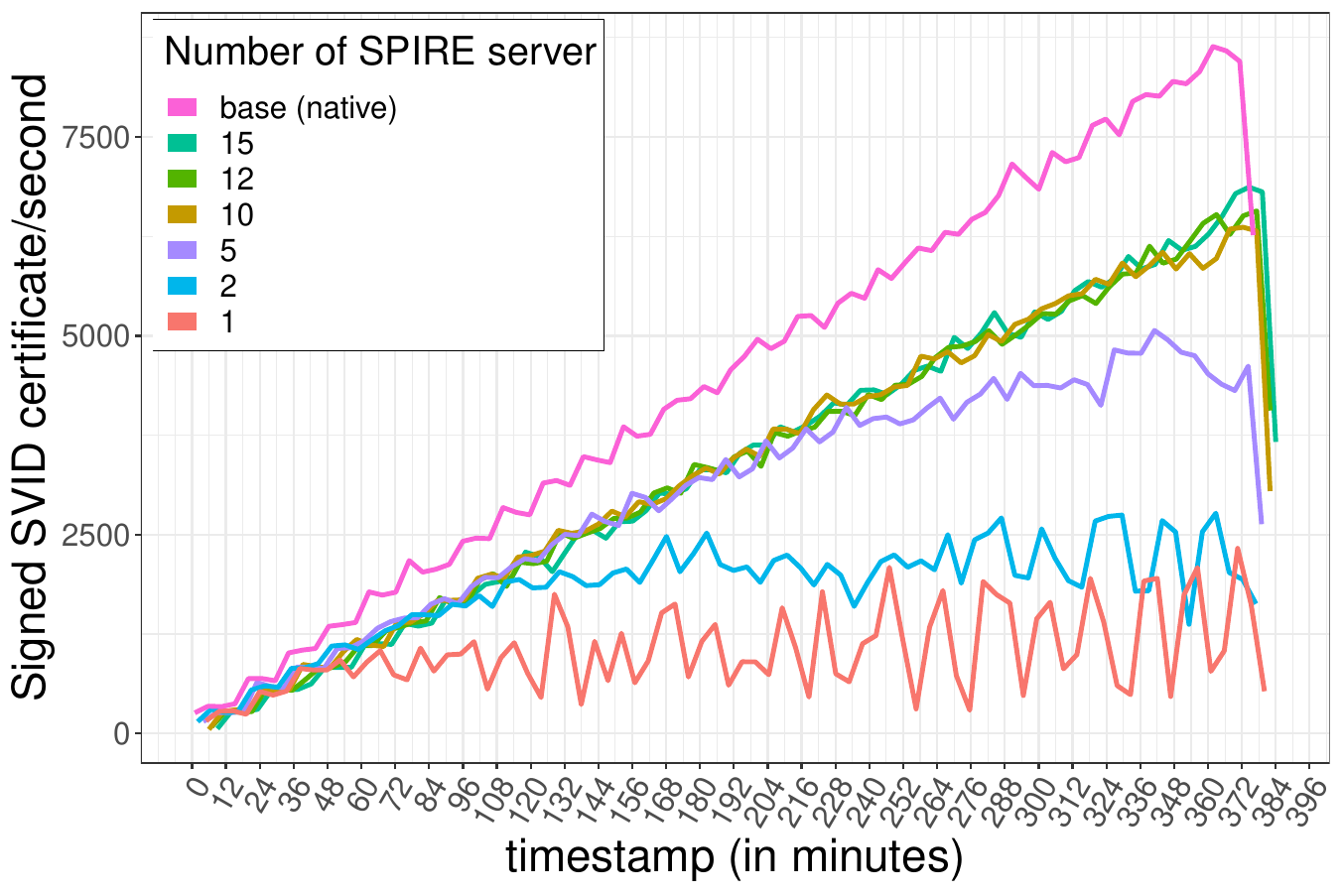}
	\caption{Certificate signing throughput on one or more trusted SPIRE servers}
	\label{fig:multiserver}
\end{figure}

For the sake of completeness, we also deployed a SPIRE server in a trusted environment. As expected, the additional overhead prevents the server from keeping up with the load, as shown in Figure~\ref{fig:multiserver} (highlighted with a red line). This is especially prominent as we do not optimize SCONE parameters on the server to reflect a simpler adoption process.

We then demonstrate that such a scenario can be mitigated horizontally and there is no bottleneck on our CRISP-equipped MariaDB. We deploy multiple TEE-equipped servers to split the load sent by the agents. These servers, deployed with an $8\ GB$ SCONE heap, share the same CRISP-enabled MariaDB backend. Similar to the previous experiment, the agents run in native mode. Figure \ref{fig:multiserver} shows the certificate signing throughput on multiple (up to 15) SPIRE servers in a trusted environment.

Our result shows that despite having a shared CRISP-based MariaDB backend, adding more SPIRE servers helps to scale the throughput on certificate signing. As a note, adding more than $10$ servers for 25 agents offers no gain, as shown in the figure. Compared with a single native SPIRE server, the average overhead with $10$ and $5$ servers is $25\%$ and $30\%$, respectively; whereas none of them is caused by CRISP. 

\section{Related Work} \label{sec:relatedwork}
\noindent{\textbf{Integrity protection through Intel SGX}.} The current work leverages existing state-of-the-art for protecting the integrity of applications running on top of Intel SGX. For example,  BROFY~\cite{Hartono2021brofy} is a multi-language compatible toolchain that can protect basic integrity, namely CPU and memory, without source code modification. LibSeal~\cite{Aublin:2018:libseal} can be used to discover integrity violations between the user and the service. Depending on the threat model, applications protected by various implementations of \textit{Shielded Execution}~\cite{graphene:2017, Arnautov:scone, Trinc:2009, orenbach2017eleos, baumann2015shielding, shinde2017panoply} should have some extent of integrity protection. Our approach extends the TEE protection to prevent storage rollback attacks. 

\smallskip
\noindent{\textbf{Protecting untrusted storage.}} Pesos~\cite{Krahn:pesos} is a policy-aware trusted object storage using Seagate Kinetic disks. Intel has released Intel PFS (Protected File System)~\cite{intel:ipfs} for use with its SGX SDK. It guarantees confidentiality and integrity but is still prone to rollback attacks. DISKSHIELD~\cite{Ahn:diskshield} extends Intel PFS and provides a defense mechanism against data tampering despite requiring SSD firmware modification. BesFS~\cite{BesFS:2020} is a library compatible with POSIX-compliant filesystem API specification. It ensures the enclave sees the last saved state, but checks are explicit. Our approach targets data CIF (Confidentiality, Integrity, and Freshness), which includes rollback resistance. Moreover, we enable protection without additional hardware besides a monotonic counter, a COTS product that is available in most servers through the TPM~\cite{ROTE:2017, Ariadne:2016}.

\smallskip
\noindent{\textbf{Preventing rollback attack.}} Our approach targets rollback attacks on top of protecting data CIF. \citeauthor{rollback:2017} provides basic rollback protection support to persistent Intel SGX applications~\cite{rollback:2017}. They implement LCM (Lightweight Collective Memory) as an alternative to a low-performance monotonic counter, similar to ROTE~\cite{ROTE:2017} and ADAM-CS~\cite{adamcs:2021}. However, all of those assume distributed settings where multiple components interact with each other through a network, adding implementation complexity. Meanwhile, CRISP applies to a single-running application with access to a monotonic counter of choice. CURE~\cite{cure:2021} requires hardware modification to enable flexible enclaves and RPMB-powered monotonic counter, of which is limited in size and adoptability. They increment the monotonic counter only at the enclave's teardown, which exposes runtime vulnerability while having a low-performance penalty. CRISP maintains the balance between protection and performance by having a tunable configuration.

PALÆMON (SCONE CAS)~\cite{Gregor:cas} has rollback protection on its server based on a monotonic counter and tag update. However, it relies on a single remote party (CAS), which only gives us consistency and freshness as long as it is available. SecureFS~\cite{Kumar:securefs} is a file system library for Intel SGX that can prevent replay attacks. It compares its approach to another Intel SGX framework, Graphene. Nimble~\cite{nimble:2023} employs replicated \textit{endorsers} and \textit{coordinators} in place of the monotonic counter as the trusted entity. Moreover, Nimble, as well as CURE~\cite{cure:2021}, also require application modification. The aforementioned approaches rely on a trusted third party to verify the local disk state. With CRISP, we delegate this task to any monotonic counter of choice, providing more flexibility.  

\smallskip
\noindent{\textbf{Confidential stateful application.}} The existence of TEE provides robust security and trust against a powerful adversary. However, primarily, it focuses more on the \textit{stateless} system~\cite{graphene:2017,Arnautov:scone}. For stateful applications, one needs to implement additional techniques to support integrity protection. Speicher~\cite{speicher:2019} is a RocksDB implementation built on Intel SGX with a data CIF guarantee. It uses a monotonic counter and encrypted storage in the form of an LSM (Log-Structured Merge Tree) data structure. EnclaveDB~\cite{enclavedb:2018}, an in-memory database server, and SGX-Log~\cite{sgxlog:2017}, which persists a secure system log, offer rollback protection by also using a monotonic counter. Despite sharing goals, previous works target their solution to a particular system. In contrast, we have no assumptions about the application. \citeauthor{spire:2022} extends SPIRE to enable attestation of Intel SGX workloads~\cite{spire:2022}; they also run SPIRE on SGX enclaves but do not consider rollback protection for the storage. The SPIRE backend stores the access control details, making it essential to protect its storage.


\section{Conclusion} \label{sec:conclusion}
This paper proposed an approach to protect stateful applications executed within TEEs. Our design leverages the SCONE runtime and its protected filesystem to embed version numbers from a monotonic counter in writing operations. The combination of Intel SGX, SCONE FSPF, and CRISP guarantees CIF (confidentiality, integrity, and freshness) of data, including protection against rollback attacks. 


Although simple, the approach combines novel mechanisms to improve the performance to acceptable levels while providing mechanisms to control the vulnerability window introduced by optimistic batching. For example, the Checker API can block and wait for the monotonic counter to commit before exposing critical information if the application cannot afford any vulnerability window.  

Finally, our evaluation includes demanding use cases with MariaDB and SPIRE. The experiments, from low-level raw disk experiments to high-complexity TPC-C benchmark, showed reasonable tradeoffs. In some cases, CRISP even outperforms native execution due to its asynchronous operations and caches.
Calling the Checker API to force pessimism can reduce performance, but if critical information is not externalized often, the impact of the checking can be small.
For the SPIRE case, the impact perceived by users and client services is almost nonexistent because of the nature of the most frequent operations (SVID certificate signings). We also demonstrated that our approach works well in distributed SPIRE settings.

\smallskip
\noindent{\textbf{Future work.}} Currently, our approach combining batching and the Checker API introduces a configurable vulnerability window. Therefore, optimizing performance still requires modifications in the application. We wish to alleviate this barrier by integrating the Checker API only at critical points. For example, integration of the Checker API with the SCONE network shield~\cite{scone:docs} enables communication to external parties to be blocked if pending increments exist. Furthermore, we intend to extend our approach in order to accommodate the possibility of breaking crashes. Specifically, we intend to adjust how FSPF is formed by applying a Write-Ahead Log (WAL), thereby enabling the reconstruction of the protected file system. Finally, this approach can be extended to other runtimes or embedded into processes within a confidential virtual machine. 

\todo{Atomic batching operations}

\section*{Acknowledgments}
We thank Gabriel P. Fernandez for the initial input on monotonic counter performance. We also thank all the reviewers for their work and helpful comments. This publication was funded by the Deutsche Forschungsgemeinschaft (DFG) as part of Germany’s Excellence Strategy -- EXC 2050/1 -- Project ID 390696704 -- Cluster of Excellence "Centre for Tactile Internet with Human-in-the-Loop" (CeTI) of Technische Universit{\"a}t Dresden, by DFG Grant 389792660 as part of TRR 248 (Foundations of Perspicuous Software Systems - CPEC), by BMBF (Federal Ministry of Education and Research) in the programme of "Souver{\"a}n. Digital. Vernetzt." on project 6G-life -- Project ID 16KISK001K, and by European Commission through the Horizon Europe Research and Innovation program under Grant Agreement No. 101092646 (CloudSkin) and No. 101092644 (NearData).

\bibliographystyle{IEEEtranN}
\small
\bibliography{reference}

\end{document}